\documentclass[a4paper,fleqn]{cas-sc}

\usepackage[numbers]{natbib}
\usepackage{setspace}
\usepackage{comment}

\def\tsc#1{\csdef{#1}{\textsc{\lowercase{#1}}\xspace}}
\tsc{WGM}
\tsc{QE}
\tsc{EP}
\tsc{PMS}
\tsc{BEC}
\tsc{DE}
\begin{document}
\let\WriteBookmarks\relax
\def\floatpagepagefraction{1}
\def\textpagefraction{.001}
\shorttitle{L{\'e}vy-noise-induced wavefront propagation}
\shortauthors{V.Semenov}

%\title [mode = title]{L{\'e}vy noise-induced wavefront propagation for bistable media, ensembles and delayed-feedback oscillators}                      

\title [mode = title]{L{\'e}vy-noise-induced wavefront propagation for bistable systems}                      

\author[1]{Vladimir V. Semenov}[orcid=0000-0002-4534-8065]
\corref{cor1}
%\ead[url]{semenov.v.v.ssu@gmail.com}
\address[1]{Institute of Physics, Saratov State University, 83 Astrakhanskaya str., 410012 Saratov, Russia}

\cortext[cor1]{Corresponding author}

\begin{abstract}
The influence of the L{\'e}vy noise's properties on wavefront propagation is analyzed on examples of ensembles of locally coupled bistable oscillators and a single bistable delayed-feedback oscillator considered as a spatially-extended system evolving in quasi-space. It is shown that additive L{\'e}vy noise allows to induce wavefront propagation in ensembles of symmetric bistable oscillators. In such a case, the direction and velocity of the noise-sustained propagation is determined both by the noise's skewness parameter and by the coupling topology (bidirectional and unidirectional coupling schemes are distinguished). In addition, additive L{\'e}vy noise induces wavefront propagation in a bistable delayed-feedback oscillator assumed to be symmetric such that its dynamics replicates the collective behaviour in the ensemble with unidirectional coupling. The wavefront propagation velocity used in this analysis is shown to be varied when adjusting the noise parameters. The revealed effects are demonstrated in the ensembles by using numerical simulation, whereas the numerical exploration of the delayed-feedback oscillator is complemented by physical experiments, showing a good correspondence and disclosing thereby the robustness of the observed phenomena.
\end{abstract}

%\begin{highlights}
%\end{highlights}


\maketitle

\doublespacing

\section{\label{sec:intro}Introduction}
The effect of wavefront propagation represents an interdisciplinary phenomenon observed in a broad spectrum of dynamical systems. In particular, models which possess the property of bistability and exhibit the phenomenon of propagating fronts and spatial domains are widely used for description of processes observed in optical systems \cite{giacomelli2012,semenov2023}, electronic circuits \cite{loecher1998,semenov2018,zakharova2025}, active metamaterials \cite{veenstra2024}, nonlinear chemical reactions \cite{schloegl1972,schloegl1983,loeber2014}, as well as for explanation of magnetic phenomena \cite{cugliandolo2010,caccioli2008}, climate and ecological changes \cite{mendez2011,ma2021,bel2012}, flame propagation \cite{zeldovich1938}, just to name a few. A manifold of effects related to wavefront propagation in bistable systems includes the phenomenon of coarsening \cite{cugliandolo2010,caccioli2008} observed in multidimensional space, where the peculiarities of front propagation are also determined by the shape of domains formed by such fronts. Coarsening is a particular form of front propagation and corresponds to the expansion of domains that invade the entire space on the cost of other domains. It is a fundamental phenomenon demonstrated in the context of physics of liquid crystals \cite{yurke1992}, physics and chemistry of materials \cite{goh2002,zhang2019,zhang2019-2,geslin2019}, physics of magnetic phenomena \cite{bray1994,cugliandolo2010,caccioli2008}, laser physics \cite{giacomelli2012,marino2014,javaloyes2015}. Besides bistable media, propagating fronts and related effects can be observed in bistable delayed-feedback oscillators \cite{giacomelli2012,semenov2018,zakharova2025} and ensembles of coupled bistable oscillators \cite{zakharova2023,semenov2023-2}.

A wide range of bistable systems exhibiting wavefront propagation determines the significance of the development of universal schemes for controlling the propagation velocity and direction. For this purpose, one can vary parameters to increase or to reduce system's asymmetry \cite{semenov2023}. In addition, stochastic control of the wavefront propagation can be applied. In particular, multiplicative noise is often used to realize the impact on the systematic part of the front dynamics \cite{engel1985,garcia-ojalvo1999,mendez2011}. In networks of coupled oscillators, one can modify the coupling topology or tune the coupling strength to control the wavefront propagation \cite{semenov2023-2,semenov2023,semenov2025_2}. 

The present paper addresses the issue of the wavefront propagation control based on the stochastic impact. In contrast to the classical approach involving multiplicative noise, an additive source of L{\'e}vy noise is introduced into the systems under study. L{\'e}vy noise is a class of stable non-Gaussian noise that exhibits long heavy tails of its distribution associated with large, potentially infinite, jumps. Various stochastic processes characterised by abrupt changes are successfully described by the L{\'e}vy noise model: spontaneous laser emission \cite{rocha2020}, non-Gaussian behavior of the heartbeat \cite{peng1993}, sensitivity of molecular motors \cite{lisowski2015}, anomalous transport in quantum-dot arrays \cite{novikov2005}, financial \cite{mantegna1999,barndorff2001} and social \cite{perc2007} processes. In certain cases, models of oscillators subject to stochastic forcing with a L{\'e}vy distribution are more suitable for description of biological neurons as compared to Gaussian noise \cite{nurzaman2011,wu2017}. In terms of noise-induced effects, L{\'e}vy noise is known to provide for controlling characteristics of noise-induced oscillations in the regime of conventional \cite{dybiec2006,dybiec2009,yonkeu2020} and self-induced \cite{yamakou2022} stochastic resonance, coherence resonance in excitable \cite{korneev2024} and non-excitable systems \cite{yonkeu2020}, noise-induced travelling waves in ensembles of excitable elements \cite{korneev2024-2} and stochastic synchronization in such systems \cite{korneev2025}. In the current paper, the mentioned list of L{\'e}vy noise-induced phenomena is extended by one more effect, L{\'e}vy noise-induced wavefront propagation. 

The wavefront propagation control is considered on two examples: ensembles of coupled bistable oscillators and a single bistable oscillator forced by delayed feedback where the delay time is much longer than the intrinsic response time of the oscillator. It is known that the presence of sufficiently long delay in single oscillators can lead to the appearance of the complex high-dimensional dynamics and provide for the observation of oscillatory regimes originally revealed and investigated  in spatially-extended systems or ensembles of coupled oscillators \cite{yanchuk2017}. In such a case, spatio-temporal phenomena are tracked down in the pure temporal dynamics of delayed-feedback oscillators by applying a spatio-temporal representation, which considers the delay interval $[0:\tau]$ in analogy with the spatial coordinate  \cite{arecchi1992,giacomelli1996}. Using this approach, the delay dynamics has been found to sustain stable chimera states \cite{larger2013,larger2015,semenov2016,brunner2018}, soliton structures \cite{garbin2015,marconi2015,romeira2016,brunner2018,semenov2018,yanchuk2019,semenov2023-3}, travelling waves \cite{klinshov2017}, and a manifold of effects associated with the property of bistability: coarsening \cite{giacomelli2012,semenov2018}, nucleation \cite{zaks2013}, deterministic and stochastic wavefront propagation and stochastic resonance \cite{zakharova2025}. Devoted to L{\'e}vy-noise-induced wavefront propagation, the current paper emphasizes the similarity of delayed-feedback oscillators and spatially-extended systems in the context of stochastic phenomena.

\section{\label{sec:medium} Ensembles}
\subsection{\label{subsec:ensemble_model} Ensemble of locally coupled oscillators and methods}

\begin{figure}[t]
\centering
\includegraphics[width=0.75\textwidth]{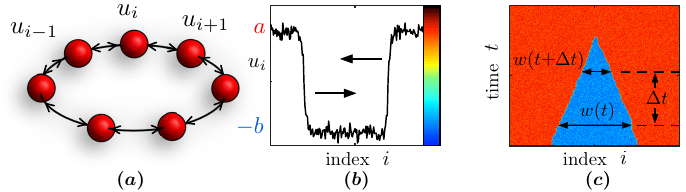}
\caption{(a) Schematic illustration of ensemble (\ref{eq:ensemble}); (b)-(c) Two propagating fronts exhibited by the model in the presence of noise (panel (b)) and the corresponding space-time plot (panel (c)). Panel (c) contains the illustration for calculating the wavefront propagation velocity $v$ in ensembles of coupled oscillators.}
\label{fig1}
\end{figure}

First, the effect of wavefront propagation is studied on an example of an ensemble of identical stochastic overdamped bistable oscillators, where the coupling is local and bidirectional [Fig.~\ref{fig1}~(a)]. The model under study is considered in the following form:
\begin{equation}
\label{eq:ensemble} 
\begin{array}{l} 
\dfrac{du_i}{dt} = -u_i(u_i-a)(u_i+b)+\xi_i(t)+\dfrac{k}{2}(u_{i+1}+u_{i-1}-2u_i), 
\end{array}
\end{equation} 
where $u_i$ are dynamical variables ($i=1,2,...,N$, where $N=200$ is the number of oscillators), $k$ is the coupling strength, $\xi_i(t)$ are statistically independent sources of noise, parameters $a,b>0$ define whether the oscillators' nonlinearity is symmetric ($a=b$) or asymmetric ($a \neq b$) which affects the wavefront propagation. In particular, the interacting oscillators are assumed to be symmetric, $a=b=1$, such that the wavefront propagation does not occur in the deterministic case, but can be induced due to the action of noise.
In more detail, symmetric case $a=b=1$ corresponds to zero front velocity in bistable media described in terms of reaction-diffusion equations, which was shown both numerically and analytically (for instance. see Refs. \cite{engel1985,loeber2014}). As emphasized below (see Sec. \ref{subsec:ensemble_results}), Eqs. (\ref{eq:ensemble}) also represent a model of a bistable medium rewritten in a lattice. Thus, zero wavefront propagation is also observed in ensemble model (\ref{eq:ensemble}) at $a=b$ in the absence of noise.

% Further analysis of the noise impact is carried out on an example of ensemble (\ref{eq:ensemble}) realizing that the same effects take place in the corresponding medium. 

Each ensemble element contains a source $\xi_i(t)$ of additive L{\'e}vy noise, which is defined as the formal derivative of the L{\'e}vy stable motion. L{\'e}vy noise is characterized by four parameters: stability index $\alpha \in (0:2]$, skewness (asymmetry) parameter $\beta\in [-1:1]$, mean value $\mu=0$ ($\mu$ is set to be zero for the strictly stable distributions \cite{janicki1994}) and scale parameter $\sigma$. Parameter $D=\sigma^{\alpha}$ is introduced as the noise intensity. If $\xi(t)$ obeys to L{\'e}vy distribution $L_{\alpha,\beta}(\xi,\sigma,\mu)$, its characteristic function takes the form \cite{janicki1994,dybiec2006,dybiec2007}:
\begin{equation}
\label{eq:characteristic_function} 
\phi(\Theta)=\int\limits_{-\infty}^{+\infty}\exp(j\Theta x)L_{\alpha,\beta}(\xi,\sigma,\mu)dx=
\left\lbrace
\begin{array}{l} 
\exp\left[-\sigma^{\alpha}|\Theta|^{\alpha}\left(1-j\beta sgn(\Theta)\tan\dfrac{\pi\alpha}{2} \right) \right],\quad \text{for } \alpha\neq 1,\\ 
\exp\left[-\sigma |\Theta|\left(1+j\beta\dfrac{2}{\pi} sgn(\Theta)\ln|\Theta| \right) \right],\quad \text{for } \alpha=1,
\end{array}
\right.
\end{equation} 
where $j$ is the imaginary unit. To generate random sequences $\xi$ corresponding to characteristic function (\ref{eq:characteristic_function}), the Janicki-Weron algorithm is used \cite{janicki1994,weron1995}: 
\begin{equation}
\label{eq:noise_generation} 
\begin{array}{l} 
\xi=\sigma S_{\alpha,\beta}\times \dfrac{\sin(\alpha(V+B_{\alpha,\beta}))}{(\cos(V))^{1/\alpha}}\times \left( \dfrac{\cos(V-\alpha(V+B_{\alpha,\beta}))}{W}\right)^{\dfrac{1-\alpha}{\alpha}}, \quad \text{for }\alpha\neq 1,\\
\xi=\dfrac{2\sigma}{\pi}\left[ \left( \dfrac{\pi}{2}+\beta V \right) \tan(V)-\beta \ln \left(\dfrac{\dfrac{\pi}{2}W\cos(V)}{\dfrac{\pi}{2}+\beta V}\right)\right], \quad \text{for }\alpha=1,
\end{array}
\end{equation} 
where $B_{\alpha,\beta}=\left( \text{arctan} \left( \beta \tan \left( \dfrac{\pi\alpha}{2}\right)\right) \right)/\alpha$, $S_{\alpha,\beta}=\left( 1+\beta^2 \tan ^2\left( \dfrac{\pi\alpha}{2}\right)\right)^{1/2\alpha}$, $V$ is a random variable uniformly distributed on $\left(-\dfrac{\pi}{2}:\dfrac{\pi}{2}\right)$, $W$ is an exponential random variable with mean 1 (variables $W$ and $V$ are statistically independent). In case $\alpha=2$, the distribution of the probability density function takes the Gaussian form with zero mean value and the variance of $2\sigma^2$. If $\alpha<2$, the distribution is non-Gaussian and the variance is infinite. Numerical simulations of Eqs.  (\ref{eq:ensemble}) are carried out by the integration by using the Heun method \cite{mannella2002} with the time step $\Delta t=10^{-4}$ or smaller. It is important to note that numerical modelling of equations including $\alpha$-stable stochastic process with finite time step implies the normalization of the noise term by $\Delta t^{1/\alpha}$ \cite{xu2016,pavlyukevich2010}.

%%%%%%%%%%%%%%%%%%fig1%%%%%%%%%%%%%%%%%%%%%%
%%%%%%%%%%%%%%%%%%%%%%%%%%%%%%%%%%%%%%%%%%

To reveal the intrinsic properties of wavefront propagation induced by L{\'e}vy noise, the evolution of space-time plots $u_i(t)$ obtained from the same initial conditions is analysed. The initial conditions are chosen to induce two wavefronts [Fig.~\ref{fig1}~(b)]: $u_i(t_0=0)=-1$ for $i \in [50:150]$ and $x_i(t_0=0)=1$ otherwise. In addition, the front propagation velocity is introduced as $v=(w(t) - w(t+\Delta t))/2\Delta t$. Here, $w(t)$ and $w(t+\Delta t)$ are the widths (integer numbers being a result of the index subtraction) of the central spatial domain corresponding to the state $x_i(t)=-1$ at the moments $t$ and  $t+\Delta t$, respectively (schematically illustrated in Fig.~\ref{fig1}~(c)). The relationship for $v$ involves the propagation of two fronts, but describes the single front propagation. For this reason, the formula for $v$ includes the factor 2. Using two fronts to calculate the propagation velocity allows to reduce inaccuracies. In summary, quantity $v$ is the velocity of the left front propagating to the right and, similarly, the velocity of the right front moving to the left [Fig.~\ref{fig1}~(b)]. The total integration time used to build the space-time plots and to calculate the wavefront propagation velocity was $t_{\text{total}}=10^4$ or larger. In the presence of noise, the wavefront propagation is characterised by fluctuations of the wavefront instantaneous position, especially in the case of ensembles subject to noise of high intensity. For this reason, the mean wavefront propagation velocity was used to analyze the stochastic ensemble dynamics. The mean wavefront propagation velocity $<v>$ was obtained as an average of 20 values of $v$ calculated on the base of 20 numerical simulations started from the same initial conditions.

\subsection{\label{subsec:ensemble_results} Results}
%%%%%%%%%%%%%%%%%%fig2%%%%%%%%%%%%%%%%%%%%%%
\begin{figure}[t]
\centering
\includegraphics[width=0.75\textwidth]{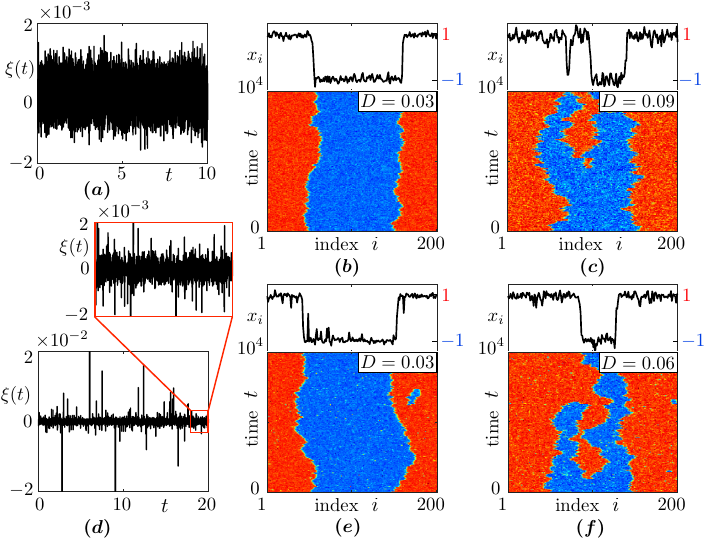}
\caption{(a)-(c) Signal $\xi(t)$ (Eqs. (\ref{eq:noise_generation})) at $\alpha=2$, $\beta=0$, $\sigma=0.01$ (a case of Gaussian noise, see panel (a)) and the evolution of wavefronts in ensemble (\ref{eq:ensemble}) subject to such stochastic impact when the noise intensity increases (panels (b) and (c)); (d)-(f) Realization $\xi(t)$ (Eqs. (\ref{eq:noise_generation})) at $\alpha=1.8$, $\beta=0$, $\sigma=0.01$ (panel (d)) and the transformation of spatio-temporal dynamics when the noise level grows (panels (e) and (f)). The ensemble parameters are: $a=b=1$, $k=1$. The color scheme for the space-time plots corresponds to the colorbar in Fig.~\ref{fig1}~(b).}
\label{fig2}
\end{figure}  
%%%%%%%%%%%%%%%%%%%%%%%%%%%%%%%%%%%%%%%%%%
In case $\alpha=2$, $\beta=0$, L{\'e}vy jumps are absent [Fig.~\ref{fig2}~(a)] and one deals with Gaussian noise sources $\xi_i(t)$. Increasing the noise intensity does not give rise to systematic wavefront propagation: the wavefront instantaneous position fluctuates, but the mean front velocity equals to zero [Fig.~\ref{fig2}~(b)]. Further growth of the noise intensity leads to destruction of spatial domains and the separating fronts. In particular, this effect is frequently observed in model (\ref{eq:ensemble}) when the noise intensity exceeds the value $D\approx0.07$ [Fig.~\ref{fig2}~(c)]. Decreasing  parameter $\alpha$ results in high-amplitude impulses in signals $\xi_i(t)$ [Fig.~\ref{fig2}~(d)]. The impulses affect the fronts such that the fluctuations of the wavefront instantaneous position are more pronounced (but the mean wavefront velocity still equals to zero) as compared to the action of Gaussian noise [Fig.~\ref{fig2}~(e)]. Moreover, the random impulses are capable of inducing new spatial domains (see the new blue domain in Fig.~\ref{fig2}~(e) which exists at $t\in [6.5\times10^3:7.5\times10^3]$) at noise intensity being not enough for producing spatial domains in the case of Gaussian stochastic force. In addition, the L{\'e}vy noise-induced destruction of the initial fronts and spatial domains occurs at lower noise intensities as compared to case $\alpha=2$ [Fig.~\ref{fig2}~(f)]. 

\begin{figure}[t]
\centering
\includegraphics[width=0.82\textwidth]{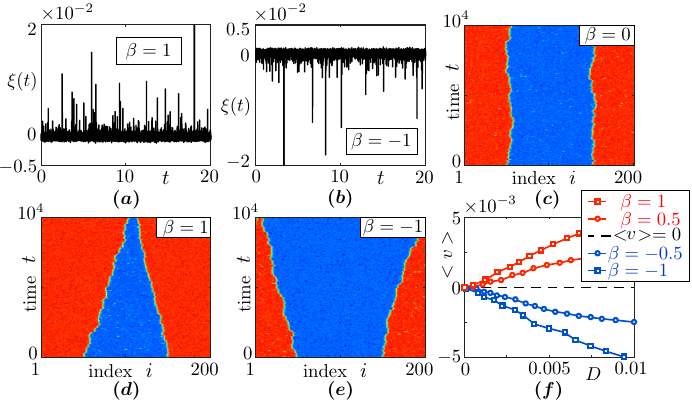}
\caption{L{\'e}vy-noise-induced wavefront propagation in ensemble (\ref{eq:ensemble}) in the presence of the noise's skewness: (a)-(b) signal $\xi(t)$ (Eqs. (\ref{eq:noise_generation}) at $\alpha=1.8$, $\sigma=0.01$) corresponding to $\beta=1$ (panel (a)) and $\beta=-1$ (panel (b)); (c)-(e) Space-time plots illustrating the spatio-temporal dynamics when varying parameter $\beta$ at fixed $\alpha=1.8$ and $D=0.01$; (f) Dependencies of the mean wavefront propagation velocity $<v>$ on the noise intensity $D$ at fixed $\alpha=1.8$ and different values of $\beta$. The ensemble parameters are the same as in the previous figure. The color scheme for the space-time plots corresponds to the colorbar in Fig.~\ref{fig1}~(b).}
\label{fig3}
\end{figure}

In case $\alpha<2$ and $\beta\neq 0$ noise signals $\xi_i(t)$ generated according to algorithm (\ref{eq:noise_generation}) become asymmetric. In particular, $\beta>0$ correspond to more pronounced jumps up  [Fig.~\ref{fig3}~(a)], whereas one observes more frequent jumps down at $\beta<0$ [Fig.~\ref{fig3}~(b)]. This affects the process of wavefront propagation in ensemble (\ref{eq:ensemble}). Indeed, zero mean wavefront propagation velocity obtained at $\beta=0$ (see the corresponding space-time plot in Fig.~\ref{fig3}~(c)) gives way to non-zero velocity values reflecting noise-sustained wavefront propagation which direction depends on the sign of the noise skewness parameter $\beta$ [Fig.~\ref{fig3}~(d),(e)]. Thus, adjusting the parameters of L{\'e}vy noise, one can tune the wavefront propagation velocity. This result is illustrated in Fig.~\ref{fig3}~(f) as dependencies of the mean wavefront propagation velocity on the noise intensity obtained at different values of parameter $\beta$.

%%%%%%%%%%%%%%%%%%fig3%%%%%%%%%%%%%%%%%%%%%%
%%%%%%%%%%%%%%%%%%%%%%%%%%%%%%%%%%%%%%%%%%

The results depicted in Fig.~\ref{fig2} and Fig.~\ref{fig3} describe the ensemble dynamics, but are also expected to be exhibited in the similar way by stochastic bistable media. To justify this statement, consider a bistable medium evolving in time and space and described by the state variable, $u=u(x,t)$. At any point $x$ in space, the dynamics is determined by a local dynamics function $f(u)=-u(u-a)(u+b)$, by an additive noise term  $\xi(x,t)$, and by a diffusion term $k_{\text{d}}\nabla^2 u$ (spatial interactions). Thus, the system equation are
\begin{equation}
\label{eq:medium} 
\begin{array}{l} 
\dfrac{du}{dt} = -u(u-a)(u+b) +\xi(x,t)+ k_{\text{d}} \nabla^2u, 
\end{array}
\end{equation} 
where $k_{\text{d}}$ is a diffusion constant. To simulate the medium in space $x$, Eq. (\ref{eq:medium}) can be rewritten in a lattice \cite{garcia-ojalvo1999} with points $x_0<x_1<x_2<...<x_N=L$ introduced such that the space cell length is $\Delta x = x_{i+1}-x_{i}$. The second-order spatial derivative can be approximated by a finite difference in order to reduce the partial differential equation to ordinary differential equations: 
\begin{equation}
\label{eq:nabla} 
\nabla^2 u(x_i,t) = \dfrac{d^2u(x_i,t)}{dx^2} \approx \dfrac{u(x_{i+1},t)-2u(x_i,t)+u(x_{i-1},t)}{\Delta x^2}.
\end{equation} 
Substituting (\ref{eq:nabla}) into (\ref{eq:medium}) and introducing $u_i(t)=u(x_{i},t)$ , $\xi_i(t)=\xi(x_{i},t)$ and $k=\dfrac{k_{\text{d}}}{\Delta x^2}$, one can rewrite Eq. (\ref{eq:medium}) and consider the bistable medium as a ring of locally coupled bistable oscillators described by Eqs. (\ref{eq:ensemble}). Thus, the same space-time plots as in Fig. \ref{fig2} and Fig. \ref{fig3} will be obtained and the same effects will observed when modelling stochastic medium (\ref{eq:medium}) and ensemble (\ref{eq:ensemble}) at the same parameter values except of the noise intensity. In more detail, the noise intensity is rescaled due to taking into account the spatial correlation, which is characterised by a new parameter of spatiotemporal noise $\xi(x,t)$, the correlation length (this aspect is described in detail in book \cite{garcia-ojalvo1999}). As a result, the dependencies of the front velocity on the noise intensity in models (\ref{eq:ensemble}) and (\ref{eq:medium}) can differ, whereas the observed phenomena are qualitatively similar, i.e. wavefronts propagate in the same direction at the same parameter sets.

\subsection{Bidirectional versus unidirectional coupling}
%%%%%%%%%%%%%%%%%%fig4%%%%%%%%%%%%%%%%%%%%%%
\begin{figure}[b]
\centering
\includegraphics[width=0.75\textwidth]{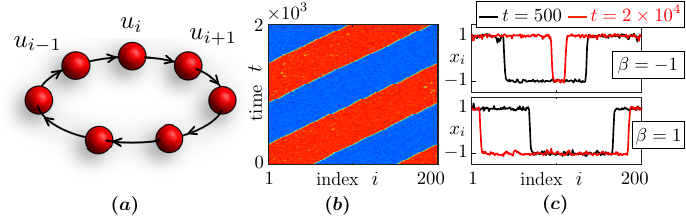}
\caption{Schematic illustration (panel (a)) and spatio-temporal dynamics (panel (b)) of unidirectionally coupled oscillators (Eqs. (\ref{eq:ensemble_unidirectional})); (c) Snapshots illustrating the instantaneous states of the ensemble at $t=500$ (coloured in black) and $t=2\times 10^4$ (coloured in red): $\beta=-1$ (upper panel) and $\beta=1$ (lower panel). Parameters are: $a=b=1$, $k=1$, $\alpha=1.8$, $D=0.01$. The color scheme for panel (b) corresponds to the colorbar in Fig.~\ref{fig1}~(b).}
\label{fig4}
\end{figure}  
%%%%%%%%%%%%%%%%%%%%%%%%%%%%%%%%%%%%%%%%%%
Next, the ensemble of stochastic bistable oscillators is considered in the presence of unidirectional coupling introduced such that 
oscillator $u_{i-1}$ affects oscillator $u_i$ (schematically illustrated in Fig.~\ref{fig4}~(a)). The ensemble model is described by the equations below:
\begin{equation}
\label{eq:ensemble_unidirectional} 
\begin{array}{l} 
\dfrac{du_i}{dt} = -u_i(u_i-a)(u_i+b)+\xi_i(t)+k(u_{i-1}-u_i).
\end{array}
\end{equation} 
Starting from the same initial conditions as in the previous section, one observes the effect wavefront propagation in model (\ref{eq:ensemble_unidirectional}): both left and right fronts propagate over the ensemble in the same direction [Fig.~\ref{fig4}~(b)]. When the nonlinearity's parameters are equal, $a=b$, the mean propagation velocities of the left and right fronts are the same. If $a \neq b$, the left and right fronts propagate to the right with different velocities such that the states $u_i(t)=a$ and $u_i(t)=b$ invade the whole space at $a>b$ and $a<b$ correspondingly.

In the context of noise-induced destruction of fronts and spatial domains registered at $\beta=0$ (see Fig.~\ref{fig2}), change of the coupling to the unidirectional configuration does not lead to the significant transformation of the dynamics as compared to the bidirectional interaction. However, the action of the L{\'e}vy noise on the process of wavefront propagation observed at nonzero $\beta$ in models (\ref{eq:ensemble}) and (\ref{eq:ensemble_unidirectional}) is characterised by the opposite result of the stochastic impact. As illustrated in Fig.~\ref{fig3}~(d),(e) on example of bidirectionally coupled oscillators, positive and negative values of $\beta$ induce expansion of states $u_i(t)=a$ and $u_i(t)=-b$ correspondingly. If the coupling is unidirectional, the situation is opposite: the state $u_i(t)=a$ invades the entire space at negative $\beta$, whereas the L{\'e}vy noise supports the state $u_i(t)=-b$ at positive $\beta$ (see Fig.~\ref{fig4}~(c)). The same result is obtained when the coupling direction is changed, i.e. when the coupling term in model (\ref{eq:ensemble_unidirectional}) is $k(u_{i+1}-u_i)$ instead of $k(u_{i-1}-u_i)$.

\section{\label{sec:delayed_oscillator} Single delayed-feedback oscillator}
\subsection{\label{subsec:delay_model} Model, experimental setup and methods}
%%%%%%%%%%%%%%%%%%fig5%%%%%%%%%%%%%%%%%%%%%%
\begin{figure}[b]
\centering
\includegraphics[width=0.95\textwidth]{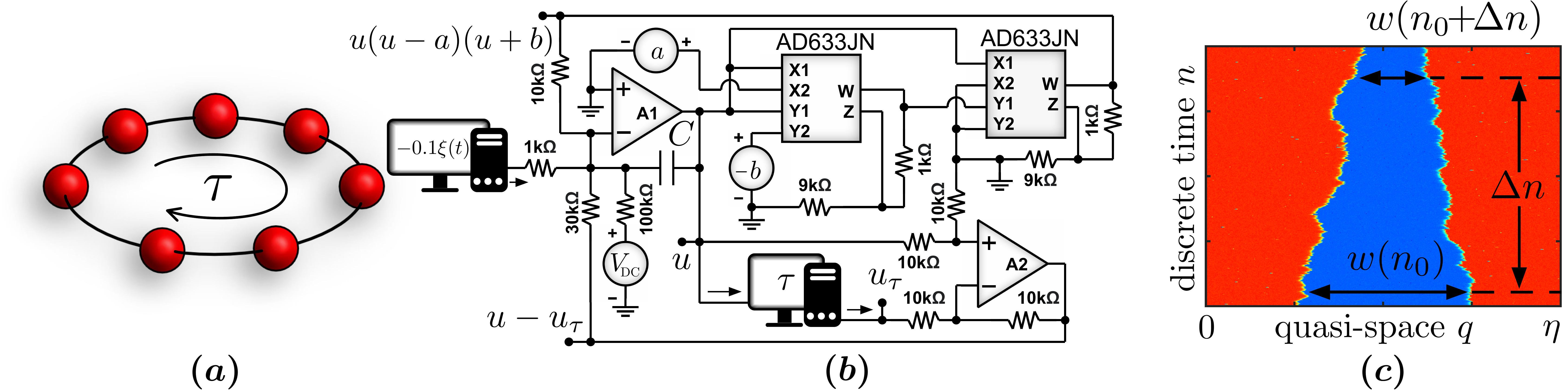}
\caption{(a) Schematic illustration of a delayed-feedback oscillator as a spatially-extended system; (b) Circuit diagram of the experimental setup (Eq. (\ref{eq:delay_experimental_model})); (c) Illustration for the calculation of the front propagation velocity $v$ in the quasi-space of a delayed-feedback oscillator.}
\label{fig5}
\end{figure}  
%%%%%%%%%%%%%%%%%%%%%%%%%%%%%%%%%%%%%%%%%%
As discussed and proved in various studies (for instance, see Ref. \cite{klinshov2017}), delay-feedback oscillators can phenomenologically describe ensembles of locally coupled oscillators with unidirectional coupling (schematically illustrated in Fig.~\ref{fig5}~(a)). Following this concept, one can expect the occurrence of the effect of L{\'e}vy-noise-induced wavefront propagation in a single delayed-feedback oscillator. Moreover, it can be supposed that the impact of the noise's asymmetry at $\beta \neq 0$ on the studied phenomenon is similar to the one observed in feed-forward ring (\ref{eq:ensemble_unidirectional}). To check the mentioned hypothesis, further exploration of the L{\'e}vy-noise-induced wavefront propagation is carried out on an example of a bistable delayed-feedback oscillator:
\begin{equation}
\label{eq:delay_numerical_model}
\begin{array}{l}
\dfrac{du}{dt}=-u(u-a)(u+b)+k (u_{\tau}-u)+\xi(t),\\
\end{array}
\end{equation}
where $u_{\tau}$ is a value of variable $u$ at the time moment $t-\tau$ ($\tau=1000$ is the delay time), parameter $k$ is the delayed-feedback strength, $\xi(t)$ is additive noise generated according to algorithm (\ref{eq:noise_generation}) and characterised by the noise intensity $D=\sigma^{\alpha}$. Similarly to ensemble models (\ref{eq:ensemble}) and (\ref{eq:ensemble_unidirectional}), parameters $a>0$ and $b>0$ are responsible for the system's symmetry and are chosen to be $a=b=1$ which corresponds to static fronts in the absence of noise. The investigations of stochastic effects observed in the single delayed-feedback oscillator are performed by means of numerical simulations and electronic experiments. Similarly to ensemble equations (\ref{eq:ensemble}) and (\ref{eq:ensemble_unidirectional}), stochastic differential equations (\ref{eq:delay_numerical_model}) are integrated using the Heun method for further time series analysis. Before the simulation starts, the initial conditions in the time range [$-\tau:0$) are specified such that two fronts are induced in the quasi-space: 
$u(t)= a$ at $ t \in \left[-\tau : -\dfrac{3}{4}\tau \right) \land \left[-\dfrac{1}{4}\tau : 0 \right)$ and $u(t)= b$ at $t \in \left[-\dfrac{3}{4}\tau : -\dfrac{1}{4}\tau \right)$.

For physical experiments, a prototype being an electronic model of system (\ref{eq:delay_numerical_model}) was developed by using the principles of analog modelling (\cite{luchinsky1998,semenov2024}). The circuit diagram of the setup is shown in Fig.~\ref{fig5}. It contains integrator A1, whose output voltage is taken as the dynamical variable $u$. The experimental facility includes a time-delay line, which was realized using personal computer complemented by an acquisition board (National Instruments NI-PCI 6251). The time delay value is constant, $\tau=27$ ms. The experimental setup is described by the equation 
\begin{equation}
\label{eq:delay_experimental_model}
\begin{array}{l}
RC\dfrac{du}{dt}=-u(u-a)(u+b)+k (u_{\tau}-u)+\xi(t)-0.1V_{\text{DC}},\\
\end{array}
\end{equation}
which includes quantities $R=10$ k$\Omega$ and $C=1$ nF being the resistance and the capacitance at integrator A1. Similarly to the mathematical model, the nonlinearity of the experimental setup is assumed to be symmetric, $a=b=1$. Parameter $k$ in Eq. (\ref{eq:delay_experimental_model}) is $k=R/30k\Omega=0.3$. Equation (\ref{eq:delay_experimental_model}) of the experimental setup can be transformed into the initial dimensionless model (\ref{eq:delay_numerical_model}) by substituting $t=t/\tau_0$  ($\tau_0=RC=0.01$~ms is the circuit's time constant) and introducing new dynamical variable $u/V_{0}$, where $V_{0}$ is the unity voltage, $V_{0}=1$~V. The experimental setup is extremely sensitive to the individual non-zero output offset voltages of the integrated circuits which inevitably present and influence the oscillators' asymmetry. Thus, the system assumed to be symmetric is in fact  asymmetric. To minimize this effect and to improve the setup, a source of low DC-voltage $V_{\text{DC}}$ (does not exceed $\pm20$mV) is included into the circuit. In more detail, values of $V_{\text{DC}}$ are tuned to achieve the slowest wavefront propagation at chosen parameters $a=b$.

The experimental facility contains a source of L{\'e}vy noise $-0.1\xi(t)$, which is implemented using the second personal computer complemented by an acquisition board NI-PCI 6251. To create a source of the L{\'e}vy noise, an algorithm of random number generation (\ref{eq:noise_generation}) is realized on the Labview platform (see book \cite{semenov2024} for more detail). The Labview program includes a block being responsible for generating random numbers $\xi$ with high frequency as an output analog signal (voltage) of the NI-board. Signal $-0.1\xi(t)$ in Fig. \ref{fig5}~(b) is assumed to be white noise, since the sample clock rate of random number generator ($f_{\text{c}}=5\times10^5$ samples per second) is much higher than the main oscillation frequency of the experimental setup. The input signal $-0.1\xi(t)$ of integrator A1 is applied to the resistor $0.1R=1$ k$\Omega$ to multiply the noise term by $10$ in Eqs. (\ref{eq:delay_experimental_model}). This provides for minimizing the amplitude limitation of the signals generated by the NI-board. In particular, the amplitude of L{\'e}vy noise jumps is much higher than the operating range of the NI-board's output voltage $[-10$ V$:10$ V]. Connection through the resistance $0.1R$ allows to realize signal $\xi(t)$, whose amplitude exceeds values in the range  $[-100:100]$. Since the experimental setup and the mathematical model have different time scales, the intensity of the experimental noise signal is introduced as $D=\sigma ^{\alpha}\left( \tau_0 f_{\text{c}}\right)^{-1/\alpha}$.

Numerically and experimentally obtained time realizations $u(t)$ are mapped onto space-time ($q,n$) by introducing $t=n\eta+q$ with an integer time variable $n$, and a pseudo-space variable $q \in [0,\eta]$, where $\eta=\tau+\varepsilon$ with a quantity $\varepsilon$, which is small as compared to $\tau$ and results from a finite internal response time of the oscillator. The procedure of estimating $\eta$ is described in detail and visualized in Ref. \cite{zakharova2025}. Generally, a unique value of the quasi-space length $\eta=\tau+\varepsilon$ is chosen in numerical and physical experiments according to the same principle: to achieve vertically oriented space-time diagrams $u(q,n)$. This corresponds to vertically symmetric diagrams such that the propagation to the left and to the right is identical as in the classical case of wavefront propagation velocity in media and ensembles. After the appropriate value of $\eta$ is found, the wavefront propagation velocity is calculated in the similar way as in the case of ensemble (\ref{eq:ensemble}) (compare Fig.~\ref{fig1}~(c) and Fig.~\ref{fig5}~(c)). In particular, the wavefront propagation is characterised by the expansion speed of the state $a$ introduced as $v=(w(n_0) - w(n_0+\Delta n))/2\Delta n$. Here, $w(n_0)$ and $w(n_0+\Delta n)$ are the widths of the central spatial domain corresponding to the state $u(t)=-1$ in the quasi-space at the moments $n_0$ and  $n_0+\Delta n$, respectively (see Fig.~\ref{fig5}~(c)). 

\subsection{\label{subsec:delay_results} Results}

\begin{figure}[t!]
\centering
\includegraphics[width=0.85\textwidth]{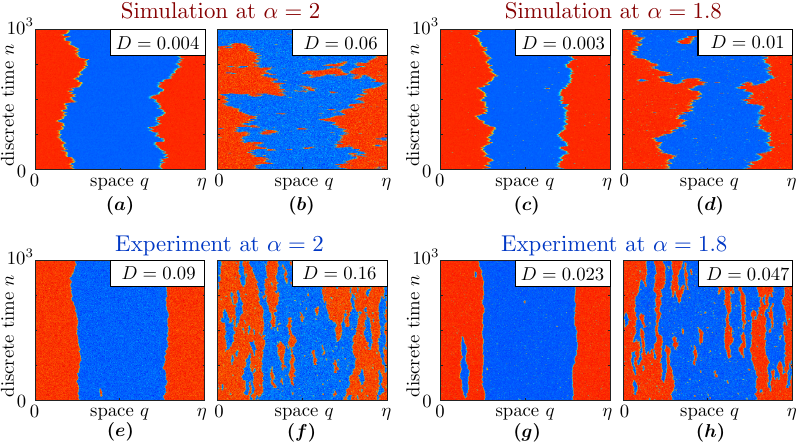}
\caption{Noise-induced destruction of wavefronts and spatial domains when increasing the noise intensity at $\alpha=2$ (panels (a),(b) and (e),(f)) and $\alpha=1.8$ (panels (c),(d) and (g),(h)). Panels (a)-(d) illustrate the effects observed in numerical experiments (see Eq. (\ref{eq:delay_numerical_model})), whereas the space-time plots in panels (e)-(h) describe the evolution of experimental setup (\ref{eq:delay_experimental_model}). Parameters are $a=b=1$, $k=0.3$, $\beta=0$, $\tau=1000$ (numerical model) and $\tau=27$ ms (experimental setup). Length of the quasi-space is $\eta=1008.146$ (panel (a)), $\eta=1006.05$ (panel (b)), $\eta=1008.123$ (panel (c)), $\eta=1007.9$ (panel (d)), $\eta=0.027214$ (panel (e)), $\eta=0.027207$ (panel (f)), $\eta=0.027219$ (panel (g)), $\eta=0.027215$ (panel (h)). The color scheme for the space-time plots corresponds to the colorbar in Fig.~\ref{fig1}~(b).
}
\label{fig6}
\end{figure}

In case $\beta=0$ the action of the L{\'e}vy noise on the wavefront propagation is the same as in ensembles of coupled oscillators and consists in destruction of fronts and domains in the quasi space. This effect is observed both in numerical and physical experiments [Fig. \ref{fig6}]. It must be noted that decreasing parameter $\alpha$ leads to lower values of the noise intensity providing for destruction of spatial structures, which was exhibited both by the numerical model and the experimental setup.

\begin{figure}[t!]
\centering
\includegraphics[width=0.75\textwidth]{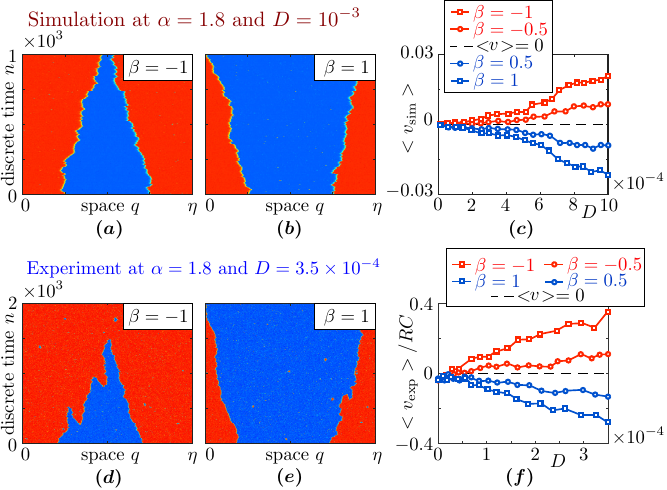}
\caption{L{\'e}vy-noise-induced wavefront propagation in a single bistable delayed-feedback oscillator (see Eqs. (\ref{eq:delay_numerical_model}) and (\ref{eq:delay_experimental_model})): numerical simulation (panels (a)-(c)) and electronic experiment (panels (d)-(f)). Panels (a),(b) and (d),(e) illustrate the evolution in the quasi-space, whereas panels (c) and (f) depict the dependence of the mean wavefront propagation velocity on the noise intensity at fixed $\alpha=1.8$ and different values of $\beta$. Other parameters are the same as in the previous figure. The quasi-space length is $\eta=1008.146$ (panels (a) and (b)), $\eta=0.027228$ (panels (d) and (e)). The color scheme for the space-time plots corresponds to the colorbar in Fig.~\ref{fig1}~(b).}
\label{fig7}
\end{figure}

In the presence of the noise's skewness, i.e. at $\beta\neq0$, the noise-sustained wavefront propagation is induced. The influence of noise on the dynamics of oscillator (\ref{eq:delay_numerical_model}) in the quasi-space is identical to the effects observed in ensemble (\ref{eq:ensemble_unidirectional}) involving unidirectional coupling. Indeed, similarly to the expansion of either spatial states in Fig.~\ref{fig4}~(c), the delay-feedback oscillator exhibits the expansion of domains corresponding to $u(t)=a$ and $u(t)=-b$ at $\beta=-1$ and $\beta=1$ correspondingly [Fig.~\ref{fig7}]. 
 
 %%%%%%%%%%%%%%%%%%fig6%%%%%%%%%%%%%%%%%%%%%%
%%%%%%%%%%%%%%%%%%%%%%%%%%%%%%%%%%%%%%%%%%
%%%%%%%%%%%%%%%%%%fig7%%%%%%%%%%%%%%%%%%%%%%
%%%%%%%%%%%%%%%%%%%%%%%%%%%%%%%%%%%%%%%%%%
 
Despite the results illustrated in Fig.~\ref{fig6} and Fig.~\ref{fig7} are supported by a physical electronic experiment which demonstrates a good qualitative agreement with the numerical simulation, there is a principal factor inevitably presenting in real bistable oscillators. Since it is not possible to realize an absolutely symmetric bistable oscillator (term $-0.1V_{\text{DC}}$ in Eq. (\ref{eq:delay_experimental_model}) allows to minimize the asymmetry, but it cannot completely eliminate its manifestation), the wavefront propagation velocity registered in physical experiments in the absence of noise signal $\xi(t)$ differs from zero. As a result, there is a certain offset in the experimentally obtained dependencies of the wavefront propagation velocities on the noise intensity as compared to the ones calculated by means of numerical simulation.

\section{\label{sec:conclusion}Conclusion}
It has been established that the properties of L{\'e}vy noise provide for inducing and controlling the wavefront propagation in ensembles of coupled bistable oscillators. It is important to note that this conclusion refers to additive noise affecting bistable oscillators with symmetric nonlinearity. In such a case, the wavefront propagation is induced and sustained due to the asymmetry of the stochastic forcing determined by the skewness parameter. The obtained result complements materials of recent paper \cite{semenov2025_2} where the ability of additive white Gaussian noise to influence the wavefront propagation is demonstrated on an example of ensembles of coupled asymmetric bistable oscillators. In summary, generalized conclusion is as follows. Additive noise can impact the process of wavefront propagation in the presence of asymmetry. The asymmetry can naturally arise (see Ref. \cite{semenov2025_2}) as well as appear due to the action of noise (for instance, by varying the skewness parameter of the L{\'e}vy noise as in the current paper). If the skewness of the L{\'e}vy noise is so small that it can be neglected, the influence of such stochastic impact consists in destruction of fronts and spatial domains when the noise's stability index decreases and the amplitude of the L{\'e}vy jumps becomes higher and higher. Thus, the L{\'e}vy noise can play both constructive (control of the wavefront propagation velocity) or destructive (noise-induced destruction of fronts) roles. 

Surprisingly, the influence of L{\'e}vy noise on the wavefront propagation depends both on the parameters of noise and on the coupling topology. In more detail, the presence on non-zero skewness parameter leads to opposite effects in ensembles with bidirectional and unidirectional coupling. This indicates that the reasons for the observed phenomena are hidden both in the intrinsic peculiarities of noise (see the asymmetric probability density functions at $\beta \neq 0$ in paper \cite{korneev2024}) and in the properties of interaction. Search for the theoretical reasons for the ability to control the wavefront propagation represents an intriguing issue for further investigations. 

The presented results cover a wide manifold of dynamical systems. In particular, it is emphasized that the phenomena realized in the ensemble with bidirectional coupling are expected to be exhibited by stochastic bistable media, whereas the collective behaviour in the presence of unidirectional coupling can be replicated in the dynamics of single delayed-feedback oscillators. The second fact is confirmed both numerically and experimentally in the current article. This is further evidence supporting the concept of a delayed-feedback oscillator as a spatially-extended system.

Results of numerical simulation and experimental study of the bistable delayed-feedback oscillator quantitatively differ from each other, which is reflected in different values of the wavefront velocities registered at the same noise intensities in the mathematical model and experimental setup. The dynamics of numerical model inevitably differs from the behaviour of the experimental setup due to a number of factors: internal fluctuations affecting the electronic circuit, different time scales, external random perturbations acting in addition to the Lévy noise source, inaccuracies taking place in all the circuit elements  (especially integrated circuits' output offsets). It is important to note that the developed Lévy noise source in not ideal. It is characterised by a finite frequency range where the power spectrum of the experimental Lévy noise signal is distributed (it is broad, up to 500 kHz, but finite). In addition, the amplitude of the experimental noise signal cannot exceed 10 Volts. As a result, Lévy jumps of high amplitudes are truncated which can impact the noise-induced dynamics. Nevertheless, numerical and experimental results are in good qualitative agreement despite the peculiarities of the experimental setup mentioned above.

\section*{Declaration of Competing Interest}
The author declares that he has no known competing financial interests or personal relationships that could have appeared to influence the work reported in this paper.

\section*{Data Availability}
The data that support the findings of this study are available from the corresponding author upon reasonable request.

\section*{Acknowledgements}
This work was supported by the Russian Science Foundation (project No. 24-72-00054).

%%% Loading bibliography style file
%%\bibliographystyle{model1-num-names}
%%%%%%%%%%%%\bibliographystyle{cas-model2-names}
%
%% Loading bibliography database
%%%%%%%%%%%%\bibliography{bibliography}

\end{document}